\documentclass[a4paper]{article}

\usepackage{INTERSPEECH_v2}
\usepackage{multirow}
\usepackage{amsmath,graphicx}
\usepackage{amsthm}
\usepackage{amssymb,amsmath}
\usepackage{bm}
\usepackage{color}

\usepackage{algorithm}
\usepackage{algorithmic}

\usepackage{cite}
\graphicspath{{D:\Surrey\Work\MyWritings\Papers\INTERSPEECH2017\figures}}

\title{Matrix of Polynomials Model based Polynomial Dictionary Learning Method for Acoustic Impulse Response Modeling}
\name{Jian Guan$^1$, Xuan Wang$^1$, Pengming Feng$^2$, Jing Dong$^3$ and Wenwu Wang$^4$}
\address{
  $^1$Harbin Institute of Technology Shenzhen Graduate School, Shenzhen, 518055, China\\
  $^2$Newcastle University, Newcastle upon Tyne, NE1 7RU, UK\\
  $^3$Nanjing Tech University, Nanjing, 211800, China\\
  $^4$University of Surrey, Guildford, GU2 7XH, UK}
\email{\{j.guan, wangxuan\}@cs.hitsz.edu.cn, p.feng2@newcastle.ac.uk, jingdong@njtech.edu.cn, w.wang@surrey.ac.uk}

\begin{document}

\maketitle
\begin{abstract}
\label{sec:abs}
We study the problem of dictionary learning for signals that can be represented as polynomials or polynomial matrices, such as convolutive signals with time delays or acoustic impulse responses. Recently, we developed a method for polynomial dictionary learning based on the fact that a polynomial matrix can be expressed as a polynomial with matrix coefficients, where the coefficient of the polynomial at each time lag is a scalar matrix. However, a polynomial matrix can be also equally represented as a matrix with polynomial elements. In this paper, we develop an alternative  method for learning a polynomial dictionary and a  sparse representation method for polynomial signal reconstruction based on this model. The proposed methods can be used directly to operate on the polynomial matrix without having to access its coefficients matrices. We demonstrate the performance of the proposed method for acoustic impulse response modeling.
\end{abstract}
\noindent\textbf{Index Terms}: polynomial dictionary learning, sparse representation, acoustic  modeling, denoising

\section{Introduction}
\label{sec:intro}
Sparse representation aims to represent a signal by the linear combination of a few atoms from an overcomplete dictionary \cite{kreutz2003dictionary,jafari2011fast}.  The dictionary can be either pre-defined using Fourier basis or wavelet basis, or adapted from training data, using dictionary learning algorithms \cite{jafari2011fast}. Dictionary learning methods usually  employ a two-step alternating optimization strategy to learn a dictionary: the first step is sparse coding, which finds the sparse representation coefficients of a signal with a given dictionary \cite{pati1993orthogonal, bruckstein2009sparse, needell2009uniform, donoho2012sparse}; and the second step is dictionary update, where the dictionary is updated to better fit the signal with the sparse representation coefficients found in the previous step \cite{aharon2006img, engan1999method, jafari2011fast, dai2012simultaneous, sadeghipoor2013dictionary}. 

Conventional dictionary learning has been studied extensively, and  used in a variety of applications, including speech denoising \cite{jafari2011fast}, and  source separation \cite{zibulevsky2001blind, yilmaz2004blind,xu2011methods}. However, it cannot be used directly to deal with the signals having time delays, such as convolutive (reverberant) signals or acoustic room impulse responses. In order to deal with such signals,  we  developed a polynomial dictionary learning technique in our recent work \cite{guan2015polynomial}, where the polynomial matrix is employed to model the signals with time lags (e.g. room impulse responses). A polynomial matrix can be represented as a polynomial with matrix coefficients (so-called polynomial of matrices model) or alternatively a matrix with polynomial elements (i.e., the matrix of polynomials model) \cite{foster2010algorithm}. In \cite{guan2015polynomial}, we  converted the proposed polynomial dictionary learning model to the conventional dictionary learning model based on the polynomial of matrices model, so that any conventional dictionary learning methods can be applied for obtaining sparse representation of the polynomial ``signal".  In \cite{guan2015polynomial}, the K-SVD algorithm\cite{aharon2006img}  and OMP algorithm  \cite{pati1993orthogonal} are employed for learning the dictionary and  for calculating the sparse representation coefficients, respectively.

In this paper, we propose a polynomial MOD algorithm (PMOD) and a  polynomial OMP algorithm (POMP) based on the matrix of polynomials model, which are the extension of the MOD algorithm \cite{engan1999method} and OMP algorithm, respectively.  Different from the method in \cite{guan2015polynomial}, the new methods can operate on the polynomial matrix directly without converting the polynomial model to a conventional model. We evaluate the proposed methods for acoustic impulse responses denoising, where the PMOD is used to learn the polynomial dictionary from a polynomial matrix modeled by acoustic impulse responses.  Both the proposed POMP algorithm and OMP algorithm are applied for the ``signal" reconstruction by using the learned dictionary.  The proposed PMOD can obtain a better performance for  acoustic signal denoising, when compared with the method in \cite{guan2015polynomial}.

The remainder of the paper is organized as follows: Section \ref{sec:background} reviews the background and  previous work about  polynomial dictionary learning; Section \ref{sec:Proposed} presents the proposed methods in details;  Section \ref{sec:simulation} shows the simulation results; and Section \ref{sec:Conclusion} gives the conclusion and potential future work.
\section{Background and previous work}
\label{sec:background}
\subsection{Polynomial matrix}
\label{polymtx}

A polynomial matrix can be expressed as either a matrix with polynomial elements, or alternatively a polynomial with matrix coefficients \cite{foster2010algorithm}. The polynomial of matrices model of a $p \times q$ polynomial matrix $\mathbf{A}(z)$ can be represented as
%\small
\begin{equation}\label{eq:1}
{\mathbf{A}(z)} = \sum\limits_{\ell=0}^{L-1} {\mathbf{A}}(\ell){z}^{-\ell}, 
\end{equation}
where ${\textbf{A}}(\ell) \in {\mathbb{C}}^{p \times q} $ is the coefficients matrix at lag $ \ell $, and $ L $ is the length of a polynomial element. The Frobenius norm (F-norm) of ${\mathbf{A}(z)}$ is defined as
%%
%\small
\begin{equation}\label{eq:2}
{\| \mathbf{A}({z}) \|_{\textit{F}}} = \sqrt{{\sum\limits_{{\mit{i}}=1}^p} {\sum\limits_{{j}=1}^q} {\sum\limits_{{\ell}=0}^{L-1}{|{a_{ij}(\ell)}|}^2}},
\end{equation}
%\normalsize
%%
where $a_{ij}(\ell)$ is the coefficient of $a_{ij}(z)$, and  $a_{ij}(z)$ is the $(i,j){\text{th}}$ polynomial element of ${\mathbf{A}(z)}$. 
\subsection{Previous work}
\label{previous}
In our previous work\cite{guan2015polynomial}, we have introduced a polynomial dictionary learning technique to deal with the signals having time delays. The aim is to learn a polynomial dictionary $ \mathbf{D}(z) \in \mathbb{R}^{n \times K}$ from the training data  $ \mathbf{Y}(z) \in \mathbb{R}^{n \times N}$, which represents  signals with time delays (such as acoustic impulse responses), so that the given ``signals"  can be represented by the learned dictionary $ \mathbf{D}(z) $. The proposed model is given as follows
%
%\small
\begin{equation}\label{eq:3}
\mathbf{Y}(\mit{z}) = {\mathbf{D}(\mit{z})}{\mathbf{X}},
\end{equation}
%\normalsize
%
where $\mathbf{X} \in \mathbb{R}^{K \times N}$ is the sparse representation coefficients.

According to \eqref{eq:1},  \eqref{eq:3} can be rewritten as 
%\small
\begin{equation}\label{eq:4}
\sum\limits_{{\ell}=0}^{{\mit{L}}-1} \mathbf{Y}(\ell){z}^{-\ell} = \sum\limits_{{\ell}=0}^{{L}-1} \mathbf{D}(\ell){z}^{-\ell}{\mathbf{X}},
\end{equation}
%\normalsize
%   
where $\mathbf{Y}(\ell) \in \mathbb{R}^{n \times N} $ and  $\mathbf{D}(\ell) \in \mathbb{R}^{n \times K}$ are the  coefficients matrices of $\mathbf{Y}(z)$ and $\mathbf{D}(z)$ at lag $\ell$, respectively. From \eqref{eq:4}, it can be seen that $\mathbf{Y}(\ell)$ can be sparsely   represented by $\mathbf{D}(\ell)$ with the same $\mathbf{X}$ for each lag $\ell$, so that \eqref{eq:3} can be converted to the conventional dictionary learning model as
%
%\small
\begin{equation}\label{eq:5}
\underline{\mathbf{Y}} = \underline{\mathbf{D}}{\mathbf{X}},
\end{equation}
%\normalsize
%%
where $ \underline{\mathbf{Y}} \in \mathbb{R}^{nL \times N} $ and the new dictionary  $\underline{\mathbf{D}} \in \mathbb{R}^{nL \times K}$ are obtained by concatenating the coefficient matrices of $\mathbf{Y}(z)$ and $\mathbf{D}(z)$ at all time lags, respectively, which are
%
%\small
\begin{equation}\label{eq:6}
 \underline{\mathbf{Y}} = \left[{\mathbf{Y}}(0);\dots;{\mathbf{Y}}(\ell);\dots;{\mathbf{Y}}(L-1)\right],
\end{equation}
%\normalsize
%
%\small
\begin{equation}\label{eq:7}
 \underline{\mathbf{D}} = \left[{\mathbf{D}}(0);\dots;{\mathbf{D}}(\ell);\dots;{\mathbf{D}}(L-1)\right].
\end{equation}
%\normalsize
%

In \cite{guan2015polynomial}, the K-SVD algorithm is employed to learn $ \underline{\mathbf{D}}$, and the OMP algorithm is used to calculate $\mathbf{X}$ for the reconstruction of $ \underline{\mathbf{Y}}$.
\section{Proposed methods}
\label{sec:Proposed}
In this section, we present methods for polynomial dictionary learning and sparse representation based on the matrix of polynomials model. More specifically, the MOD algorithm and the OMP algorithm are extended to the polynomial cases,  for dictionary update and sparse approximation,  respectively. 

\subsection{Polynomial MOD}
\label{sec:pmod}
The proposed polynomial dictionary learning method updates the dictionary by optimizing the following cost
\begin{equation}\label{eq:8}
{\mathbf{D}(z)}^{(n+1)} = \underset{\mathbf{D}(z)}{\text{argmin}}{\|\mathbf{Y}(z) - {\mathbf{D}(z)}{\mathbf{X}}^{(n)} \|_F^2},
\end{equation}
where $\mathbf{X}^{(n)}$ is the sparse representation  matrix at the $n\text{th}$ iteration. Here,  \eqref{eq:8} can be seen as a polynomial least-squares problem, so that the same strategy as in MOD can be  employed to solve \eqref{eq:8}, where the dictionary can be updated by calculating the least-squares solution of \eqref{eq:8}, which is
\begin{equation}\label{eq:9}
{\mathbf{D}(z)}^{(n+1)} = \mathbf{Y}(z){{\mathbf{X}}^{(n)}}^T{({\mathbf{X}}^{(n)}{{\mathbf{X}}^{(n)}}^T)}^{-1}.
\end{equation}

By using the polynomial least-squares, the proposed polynomial MOD method can update the dictionary directly without operating on the coefficient matrices as in our previous work  \cite{guan2015polynomial}. The proposed PMOD algorithm is given in Algorithm \ref{alg:pmod}, and a polynomial sparse representation method is discussed next.

%%%% Algorithm 1
\begin{algorithm}
\caption{Polynomial MOD (PMOD)}\label{alg:pmod}
\begin{algorithmic}
\STATE $\mathbf{Input}$: $\mathbf{Y}(z)$, $I_n$
\STATE $\mathbf{Initialization}$: $\mathbf{D}(z)^{(0)} = \mathbf{Y}(z)(:,1:K) $, $I_{n}=80$.
\STATE $\mathbf{Iterations}$:
\FOR {$i = 1,\cdots , I_{n}$}
\STATE $\mathbf{Sparse\ coding}$:
\STATE Calculating $\mathbf{X}$ by using OMP as in \cite{guan2015polynomial}.
% or polynomial OMP in Algorithm \ref{alg:pomp}.
\STATE $\mathbf{Polynomial\ dictionary\ update}$: 
\STATE Updating $\mathbf{D}(z)$ by using \eqref{eq:9}.
\ENDFOR
\STATE $\mathbf{Output}$: $\mathbf{D}(z)$ and $\mathbf{X}$
\end{algorithmic}
\end{algorithm}
%%%
%%

%
\subsection{Polynomial OMP}
\label{sec:sparse representation}
As the aim of our work is to deal with the signals with time delays, once the polynomial dictionary $\mathbf{D}(z)$ is learned by our
proposed methods,  the sparse representation coefficients matrix $\mathbf{X}$ also needs to be calculated for the reconstruction of the polynomial matrix $\mathbf{Y}(z)$. Here, we present a polynomial sparse representation method  by extending the OMP algorithm \cite{pati1993orthogonal} to the polynomial case. 

The polynomial sparse representation coefficients can be calculated by optimizing the following cost
\begin{equation}\label{eq:10}
\begin{aligned}
& \underset{\mathbf{x}}{\text{min}}{\|\mathbf{y}(z) - {\mathbf{D}(z)}{\mathbf{x}} \|_F^2} \\
& \text{subject to} {\ }\|\mathbf{x}\|_0 \le K_{max},\\
\end{aligned}
\end{equation}
%\normalsize
where  the ``signal" $\mathbf{y}(z)$  is a column of $\mathbf{Y}(z)$, and  $K_{max}$ is the max number of non-zero elements in $\mathbf{x}$.

The proposed POMP algorithm employs the same strategy as the OMP algorithm to calculate the sparse representation coefficients $\mathbf{x}$ to approximate the ``signal" $\mathbf{y}(z)$. The OMP algorithm calculates the sparse representation coefficients by iteratively selecting the best-matched atoms 
from the dictionary to approximate the signal, where the best-matched atom selected is the one that is most correlated with the
residual at each iteration. However, the similarity measure between two polynomial vectors (e.g. polynomial residual and polynomial dictionary atom) cannot be directly achieved with the inner product between the residual and the atoms as in the original OMP.

In our proposed method, we calculate the distance between the polynomial residual and polynomial dictionary atom by using the F-norm as the similarity measure, so that the best-matched atom is the one that has the smallest F-norm error with the polynomial residual at each iteration, which can be formulated as follows
\begin{equation}\label{eq:11}
{k_0}= \underset{k}{\text{argmin}}{\| \mathbf{d}_k(z)-\mathbf{r}{(z)}^{(j-1)}\|_F^2},\ k = 1, \cdots, K,
\end{equation}
where $\mathbf{d}_{k}(z)$ is an atom in $\mathbf{D}(z)$, $\mathbf{r}{(z)}^{(j-1)}$ is the polynomial residual vector obtained at the $(j-1){\text{th}}$ iteration, and ${k_0}$ is the index of the selected atom at the $j{\text{th}}$ iteration. Then, ${k_0}$ is added to the support set $S^{(j)}$, and the sub-dictionary $\mathbf{D}_{S^{(j)}}(z)$  is also updated by adding the selected atom $\mathbf{d}_{k_0}(z)$. Then, the representation coefficients  can be updated by minimizing the following cost function 
\begin{equation}\label{eq:12}
\begin{aligned}
&{\underset{\mathbf{x}}{\text{min}}{\|\mathbf{y}(z) - {\mathbf{D}_{S^{(j)}}(z)}{\mathbf{x}}}} \|_F^2 \\
& \text{subject to} {\ }\|\mathbf{x}\|_0 \le K_{max},\\
\end{aligned}
\end{equation}
where \eqref{eq:12} is a polynomial least-squares problem. Note that, according to \eqref{eq:4} and \eqref{eq:5}, the coefficients $\mathbf{x}$ should satisfy the linear combination between $\mathbf{y}(z)$ and ${\mathbf{D}_{S^{(j)}}(z)}$ at all lags. So that, the coefficient can be updated by the least-squares solution of \eqref{eq:12}, which is
%
%\small
\begin{equation}\label{eq:13}
\mathbf{x}^{(j)}= {({{\underline{\mathbf{D}}^{T}_{{S}^{(j)}}}{\underline{\mathbf{D}}_{{S}^{(j)}}}})}^{-1}{\underline{\mathbf{D}}^{T}_{{S}^{(j)}}}{\underline{\mathbf{y}}},
\end{equation}
%\normalsize
where  ${\underline{\mathbf{y}}}$ and $\mathbf{D}_{S^{(j)}}(z)$ are constructed according to \eqref{eq:6} and \eqref{eq:7}. Then, the polynomial residual can be updated as
%
%
%\small
\begin{equation}\label{eq:14}
\mathbf{r}^{(j)}(z) = \mathbf{y}(z) - \mathbf{D}_{S^{(j)}}(z){\mathbf{x}^{(j)}}.
\end{equation}
%\normalsize
%
The proposed POMP algorithm is summarized in Algorithm  \ref{alg:pomp}.
%
%%%% Algorithm 2
\begin{algorithm}
\caption{Polynomial OMP (POMP)}\label{alg:pomp}
\begin{algorithmic}
\STATE {\textbf{Input:}}  $\mathbf{y}(z)$,  $\mathbf{D}(z)$,  $K_{max}$
\STATE $\mathbf{Initialization}$: residual $\mathbf{r}(z)^{(0)}  = \mathbf{y}(z)$,  $\mathbf{x} = 0$,   $S^{0}=\emptyset$, $\epsilon = 10^{-6}$.
\STATE $\mathbf{Iteration}$:
\FOR {$j = 1,\ldots , K_{max}$}
\STATE $\mathbf{Atom\ selection}$: 
\STATE $k_0 = \underset{k}{\text{argmin}}{\| \mathbf{d}_k(z)-\mathbf{r}(z)^{(j-1)}\|_F^2}$,{\ } $k = 1, \cdots, K$
%Compute F-norm $\| \mathbf{d}_k(z)-r(z)^{(j-1)}\|_F^2$, where $\mathbf{d}_k(z) \in \mathbf{D}(z)$, $k = 1, \cdots, K$, and $k_0 = \underset{k}{\text{argmin}}{\| \mathbf{d}_k(z)-r(z)^{(j-1)}\|_F^2}$.
\STATE $\mathbf{Support\ set\ update}$: 
$S^{(j)} = S^{(j-1)} \bigcup \{k_0\}$
%$\mathbf{D}_{k_0}(z)^{(j)} =\{\mathbf{D}_{k_0}(z)^{(j-1)},\mathbf{d}_{k_0}(z)\}$.
\STATE $\mathbf{Coefficient\ update}$: %\small
$\mathbf{x}^{(j)}= {({{\underline{\mathbf{D}}^{T}_{{S}^{(j)}}}{\underline{\mathbf{D}}_{{S}^{(j)}}}})}^{-1}{\underline{\mathbf{D}}^{T}_{{S}^{(j)}}}{\underline{\mathbf{y}}}$
\STATE $\mathbf{Residual\ update}$: $\mathbf{r}(z)^{(j)} = \mathbf{y}(z) - \mathbf{D}_{S^{(j)}}(z){\mathbf{x}^{(j)}}$
\STATE $\mathbf{Stopping\ rule}$: If $\|{\mathbf{r}(z)}^{(j)}\|_F^2 \le \epsilon$, then $\mathbf{x}_{opt}=\mathbf{x}^{(j)}$, and break, else continue.
\ENDFOR
\STATE {\textbf{Output:}} $\mathbf{x}_{opt}$
\end{algorithmic}
\end{algorithm}
%%
%%
%%%%
\section{Simulations and results}
\label{sec:simulation}
In this section, we apply our proposed polynomial dictionary learning method to represent the signals with time delays, e.g. acoustic impulse responeses. Polynomial dictionaries are learned from an acoustic impulse responses modeled polynomial matrix. The learned dictionaries are then  used to recover the noise corrupted acoustic impulse responses, where the method in \cite{guan2015polynomial} is employed as a baseline for learning the polynomial dictionaries for performance comparison.  The performance is measured by the polynomial ``signal" reconstruction error, which is defined as 
\begin{equation}\label{eq:15}
\textit{R}_{err} =\frac{{\|{\mathbf{Y}({z})- \hat{\mathbf{Y}}({z})}\|}_{{F}}^{2}}{{\| \mathbf{Y}({z})\|}_{{F}}^{2}},
\end{equation}
where  $\mathbf{Y}({z})$ is the original signal and   $\hat{\mathbf{Y}}({z})$ is the reconstructed signal.

\subsection{Acoustic impulse responses generation and modeling}
\label{sec:database}
The proposed method is evaluated on acoustic impulse response signals generated by the room image model \cite{allen1979image} as in \cite{guan2015polynomial}. A total number of 1000 impulse responses are used as training signals, the length of each impulse response is 14400. By applying our proposed method, a $ 10 \times 72000$ polynomial matrix with $20$ lags is designed to model the training acoustic signals, where each element of the polynomial matrix has 20 lags, which can be seen as a finite impulse response (FIR) with a  length of 20 samples. Each test acoustic signal is split into $720$ segments, where the length of each segment is set to be  20, so that the test signal can be modeled by a $10 \times 72$ polynomial matrix with 20 lags.
\subsection{Results and analysis}
\label{sec:results}
First, we carried out an experiment to compare the performance between the proposed method and method in \cite{guan2015polynomial} for acoustic signal denoising, where the polynomial dictionaries were trained by PMOD and the method in \cite{guan2015polynomial}, respectively. The number of iterations for training the dictionaries in both PMOD and the method in \cite{guan2015polynomial} were set as 80. The size of the dictionaries were designed to be the same, which was $10 \times 400$ with $20$ lags. 
The learned dictionaries were used to recover the noise corrupted acoustic signal, where white Gaussian noise with zero-mean and unit variance, set at different levels of signal-to-noise (SNR) were added  to the test signal. By applying the PMOD trained dictionary, both OMP and POMP were used to calculate the sparse representation coefficients for acoustic signal reconstruction, which were denoted by PMOD+OMP and PMOD+POMP, respectively. The sparsity for training the dictionaries and calculating the sparse representation coefficients was all set to be 3.  The experiments were conducted for 20 realizations at each noise level,  and the average reconstruction errors were given in Table \ref{tab:1}.

\begin{table}[htb]
\centering
\caption{Performance comparison in terms of Reconstruction Error ($ \times 10^{-2}$)  between the proposed method and the  baseline method in \cite{guan2015polynomial} for acoustic signal denoising at different noise levels.}
\label{tab:1}
\begin{tabular}{@{}cccccc@{}}
%\begin{tabular}{l@{}ccccc@{}}
%\begin{tabular}{r@{\ }lrrrr}
\toprule
{\textbf{Noise levels (dB)}} & $-10$    & $0$     & $10$    & $20$    & $30$    \\ \midrule
{\textbf{Method in \cite{guan2015polynomial} }}         & $249.71$ & $37.19$ & $15.62$ & $15.45$ & $15.43$ \\
{\textbf{PMOD+OMP}}      & $248.92$ & $37.05$ & $15.23$ & $15.07$ & $15.05$ \\
{\textbf{PMOD+POMP}}      & $228.63$ & $37.26$ & $20.27$ & $20.16$ & $20.15$ \\ \bottomrule
\end{tabular}
\end{table}

From Table \ref{tab:1}, we can see that the PMOD+OMP can obtain better performance than the other two methods for acoustic signal recovery when the noise is at the level from 0 dB to 30 dB, whereas the PMOD+POMP gives better recovery accuracy for SNR at -10 dB. We found that  a larger dictionary tends to give better reconstruction accuracy for most noise levels tested by using all these methods when using the training polynomial matrices with same lags (e.g. the polynomial matrices modeled with 20 lags). Therefore,   dictionaries with the size of $10 \times 400$ were trained. The proposed methods can achieve better reconstruction accuracy in this situation. 

\begin{figure}[htb]
\begin{minipage}[b]{1.0\linewidth}
  \centering
  \centerline{\includegraphics[width=8cm, height=2cm]{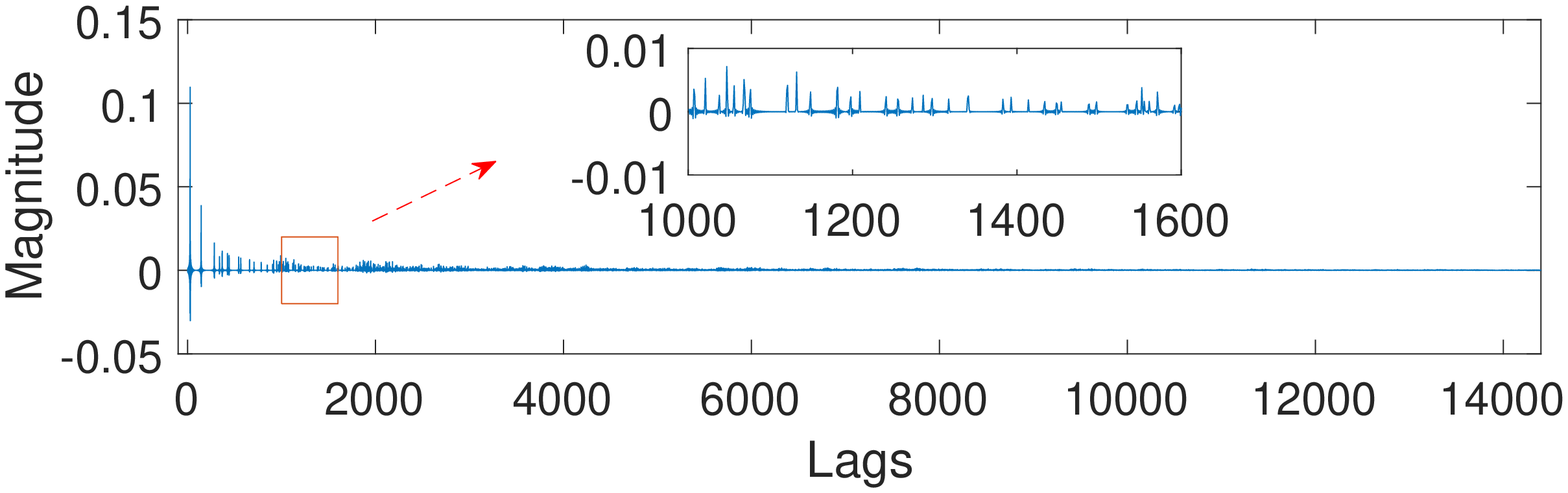}}
%\vspace{-5mm}
  \centerline{(a)}\medskip
\end{minipage}
\begin{minipage}[b]{1.0\linewidth}
  \centering
  \centerline{\includegraphics[width=8cm, height=2cm]{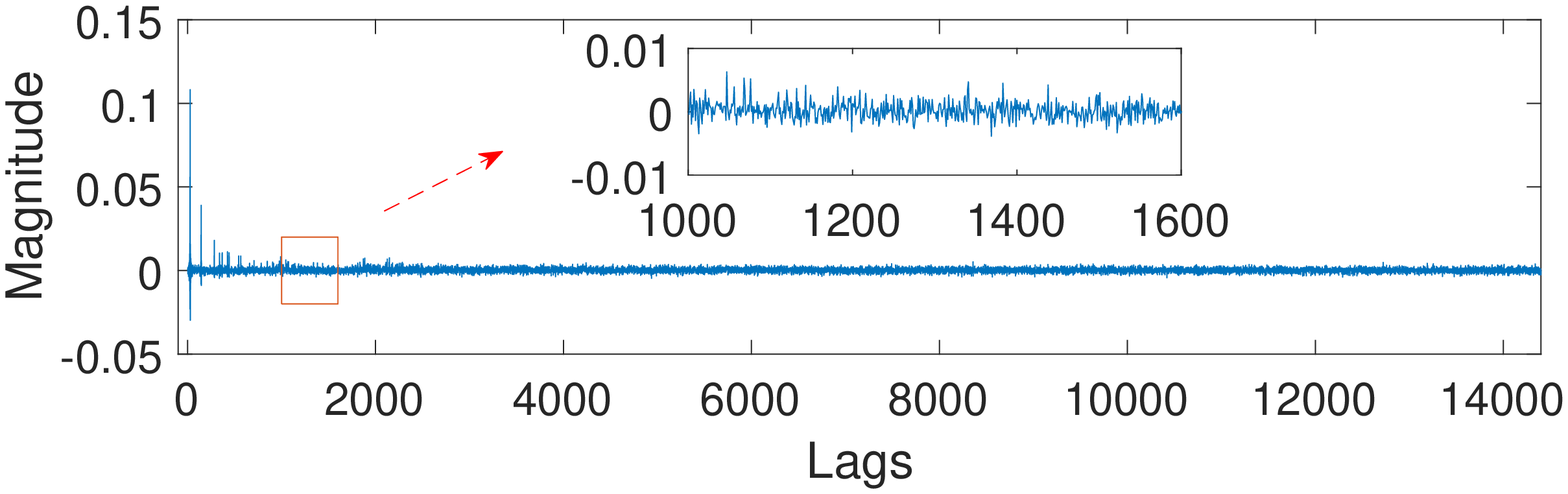}}
%  \vspace{1.5cm}
  \centerline{(b)}\medskip
\end{minipage}
\hfill
\begin{minipage}[b]{1.0\linewidth}
  \centering
  \centerline{\includegraphics[width=8cm, height=2cm]{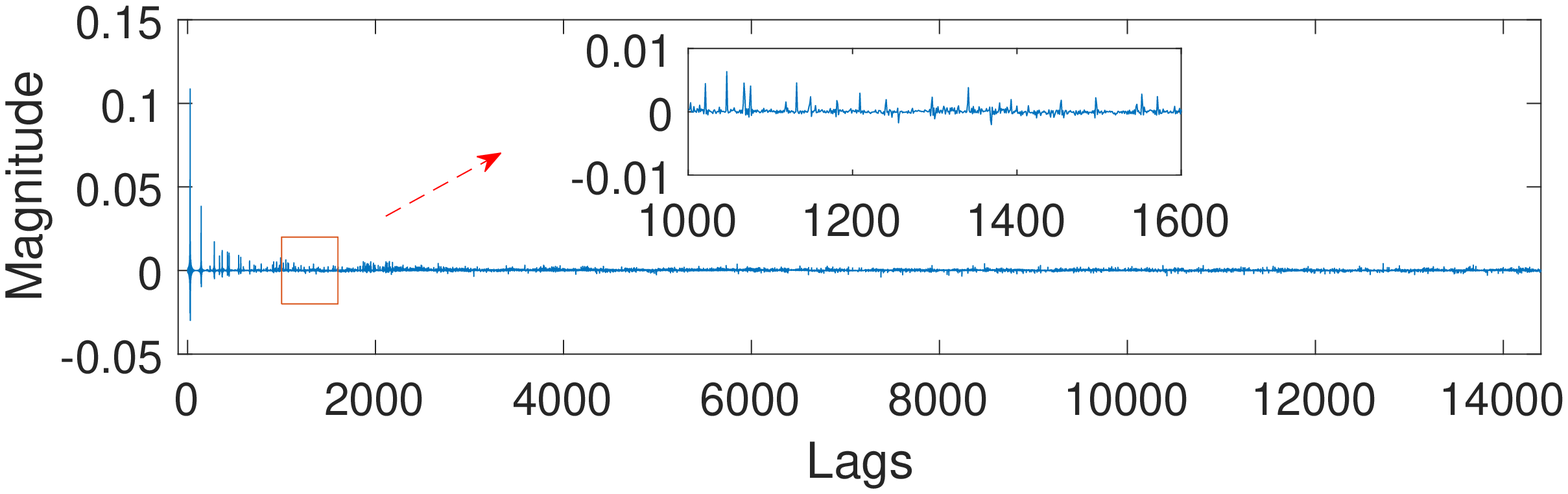}}
%  \vspace{1.5cm}
  \centerline{(c)}\medskip
\end{minipage}
\begin{minipage}[b]{1.0\linewidth}
  \centering
  \centerline{\includegraphics[width=8cm, height=2cm]{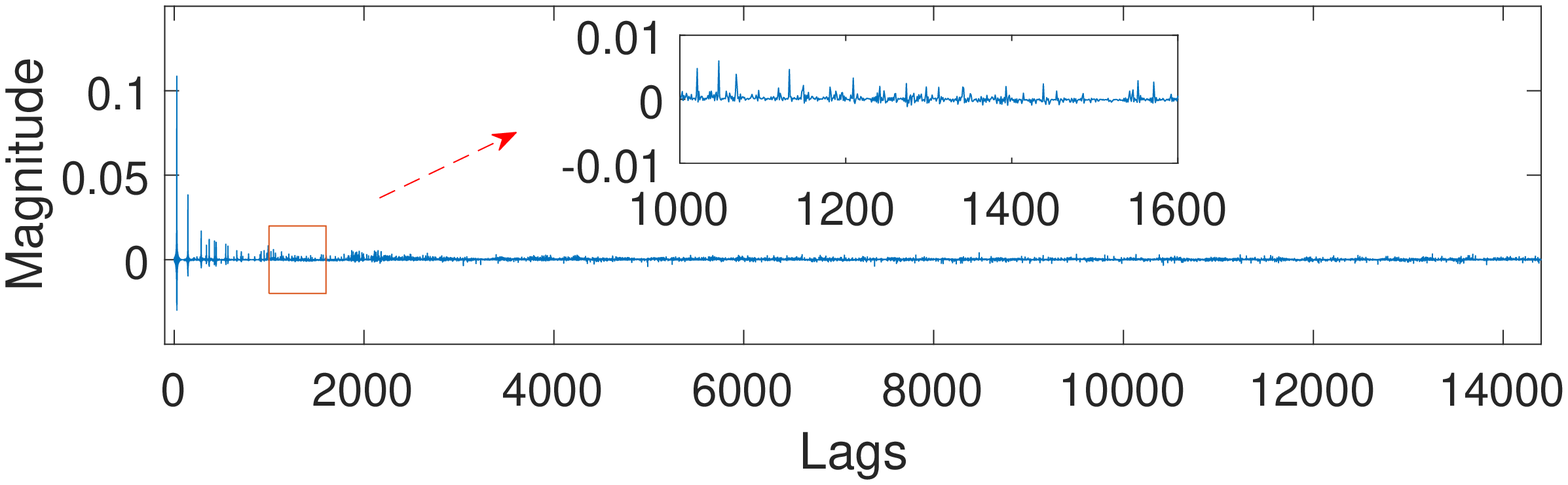}}
%  \vspace{1.5cm}
  \centerline{(d)}\medskip
\end{minipage}
\begin{minipage}[b]{1.0\linewidth}
  \centering
  \centerline{\includegraphics[width=8cm, height=2cm]{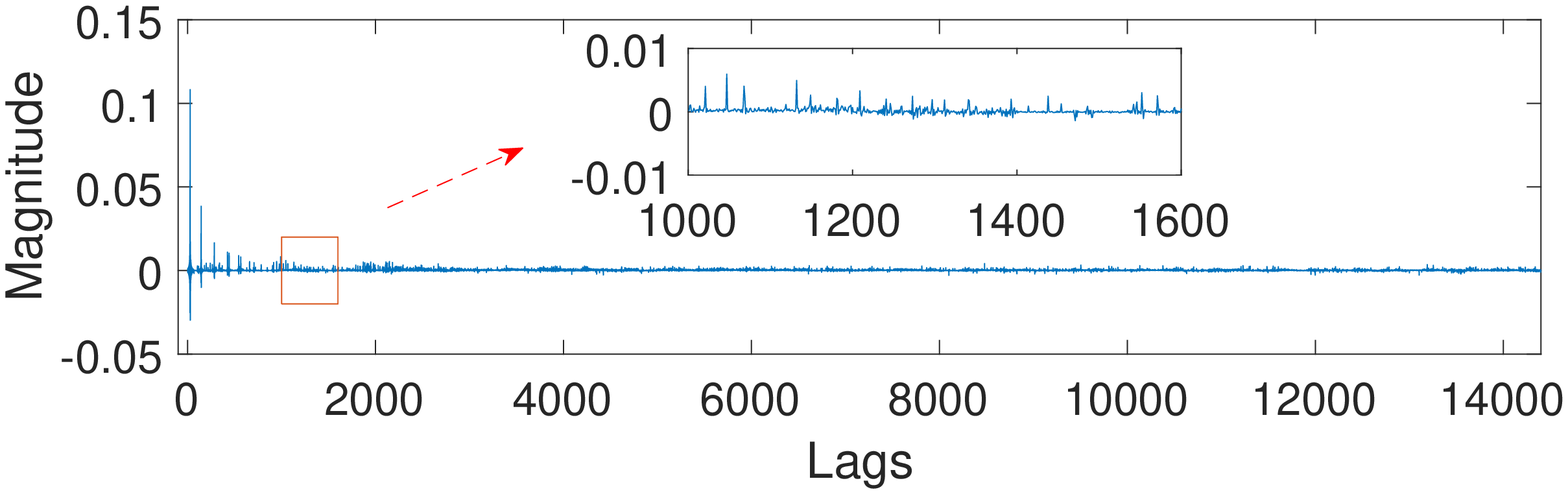}}
%  \vspace{1.5cm}
  \centerline{(e)}\medskip
\end{minipage}
\caption{Illustration of the polynomial dictionary learning methods for noise corrupted acoustic impulse reconstruction:  Clean acoustic impulse response (a);  Noisy acoustic impulse response (b);  Reconstructed acoustic impulse response by the method in \cite{guan2015polynomial}  (c);  by PMOD+OMP  (d); and by PMOD+POMP  (e).}
\label{fig:1}
\end{figure}

Then, another experiment was conducted to give an illustration of the acoustic signal denoising. The size of the learned dictionaries was  the same as that in the previous experiment, and 5 dB white Gaussian noise was added to the test signal. Figure \ref{fig:1} shows the result of the polynomial dictionary learning methods for noisy acoustic signal reconstruction, using the  method in \cite{guan2015polynomial}, PMOD+OMP, and PMOD+POMP. From Figure \ref{fig:1}, we can see that the denoising performance of the polynomial dictionary learning methods used are quite similar, all these methods can recover the source signal to a certain degree.

%%%%

\begin{figure}[htb]
\begin{minipage}[b]{1.0\linewidth}
  \centering
  \centerline{\includegraphics[width=8cm, height=2cm]{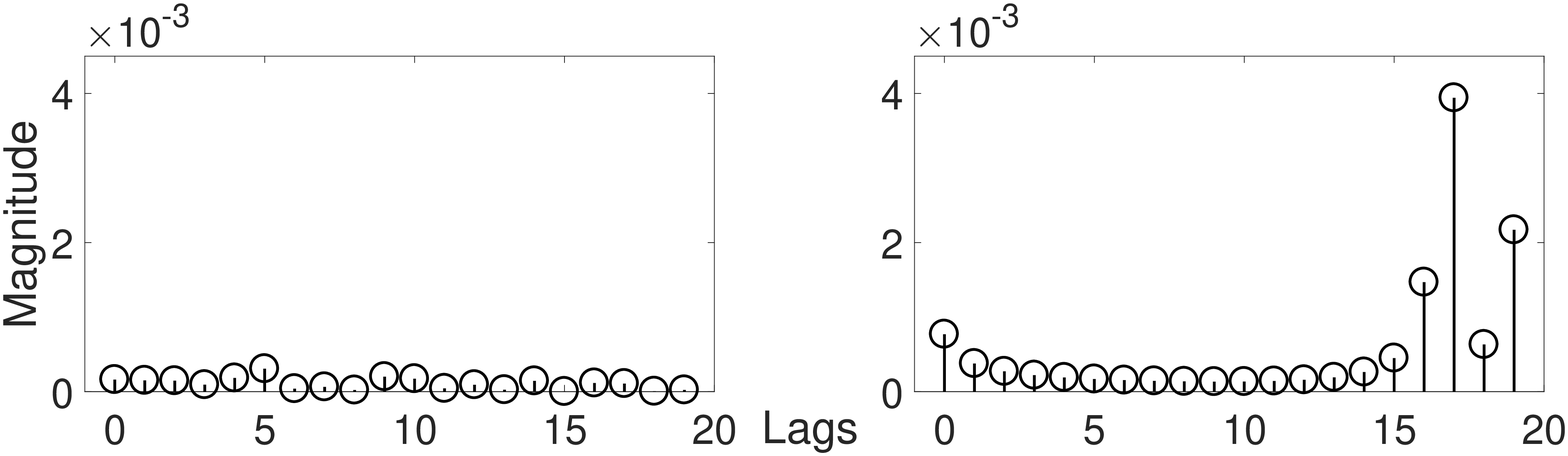}}
%\vspace{-5mm}
  \centerline{(a)}\medskip
\end{minipage}
\begin{minipage}[b]{1.0\linewidth}
  \centering
  \centerline{\includegraphics[width=8cm, height=2cm]{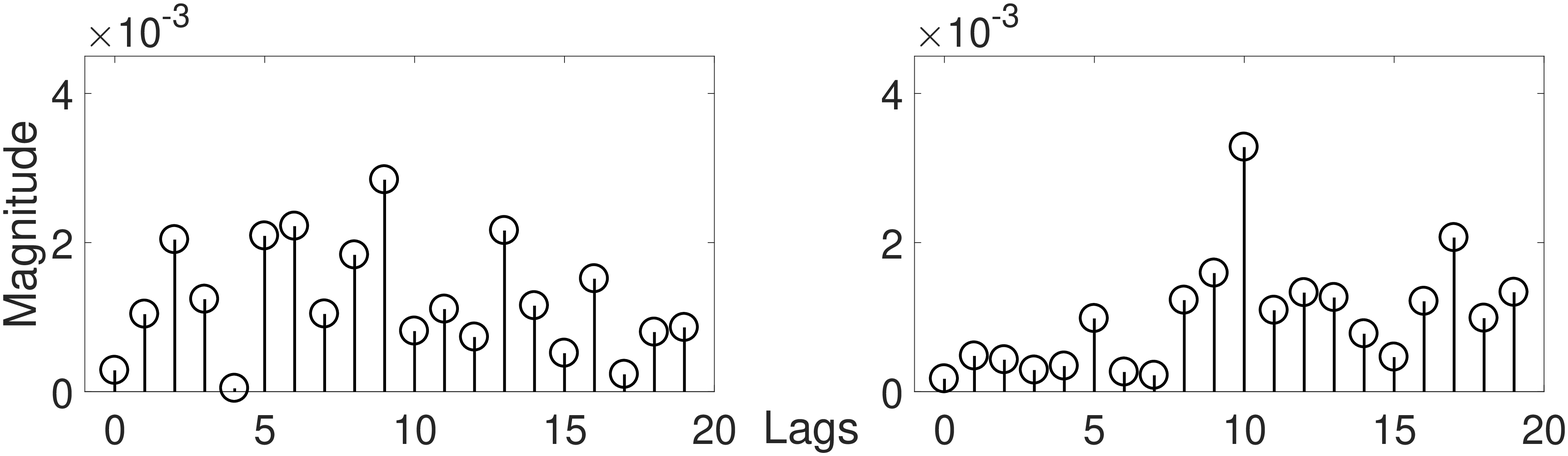}}
%  \vspace{1.5cm}
  \centerline{(b)}\medskip
\end{minipage}
\hfill
\begin{minipage}[b]{1.0\linewidth}
  \centering
  \centerline{\includegraphics[width=8cm, height=2cm]{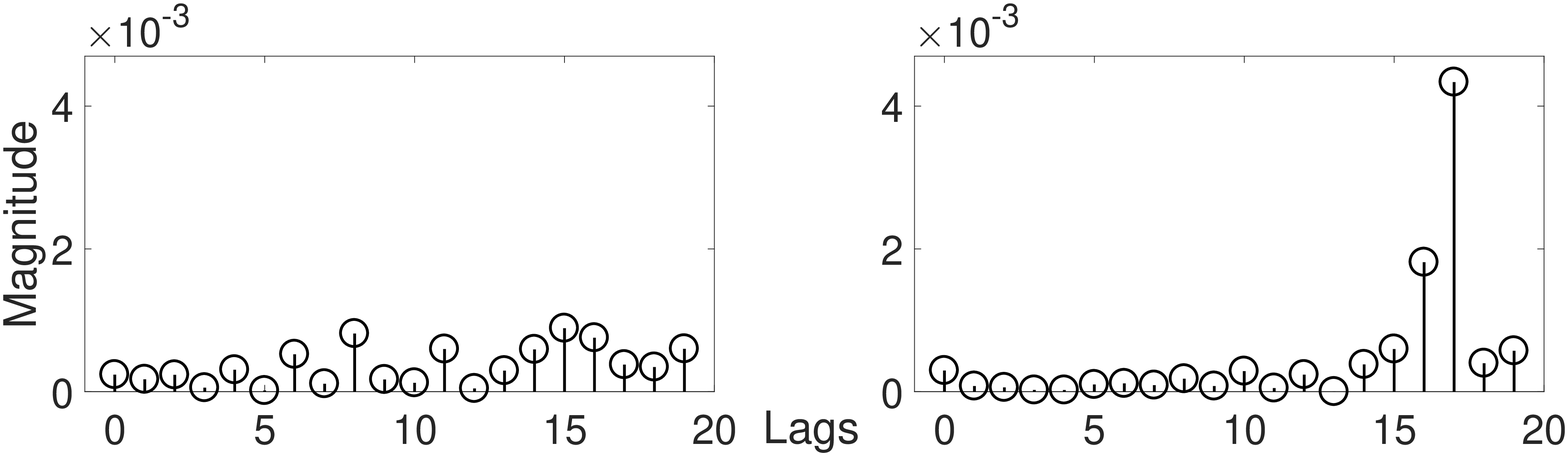}}
%  \vspace{1.5cm}
  \centerline{(c)}\medskip
\end{minipage}
\begin{minipage}[b]{1.0\linewidth}
  \centering
  \centerline{\includegraphics[width=8cm, height=2cm]{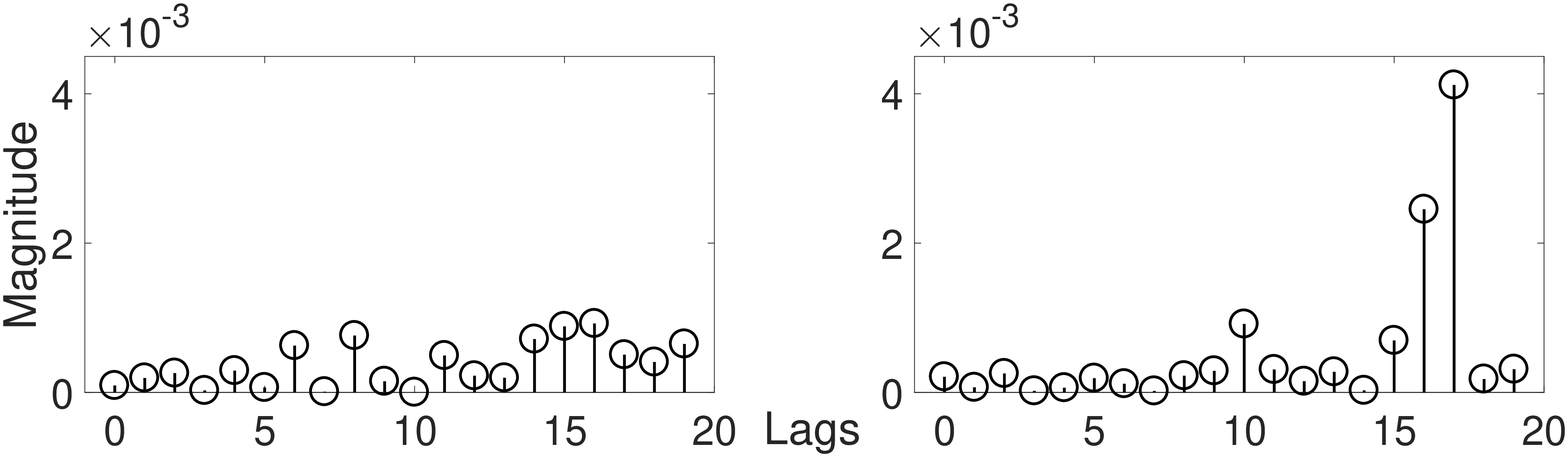}}
%  \vspace{1.5cm}
  \centerline{(d)}\medskip
\end{minipage}
\begin{minipage}[b]{1.0\linewidth}
  \centering
  \centerline{\includegraphics[width=8cm, height=2cm]{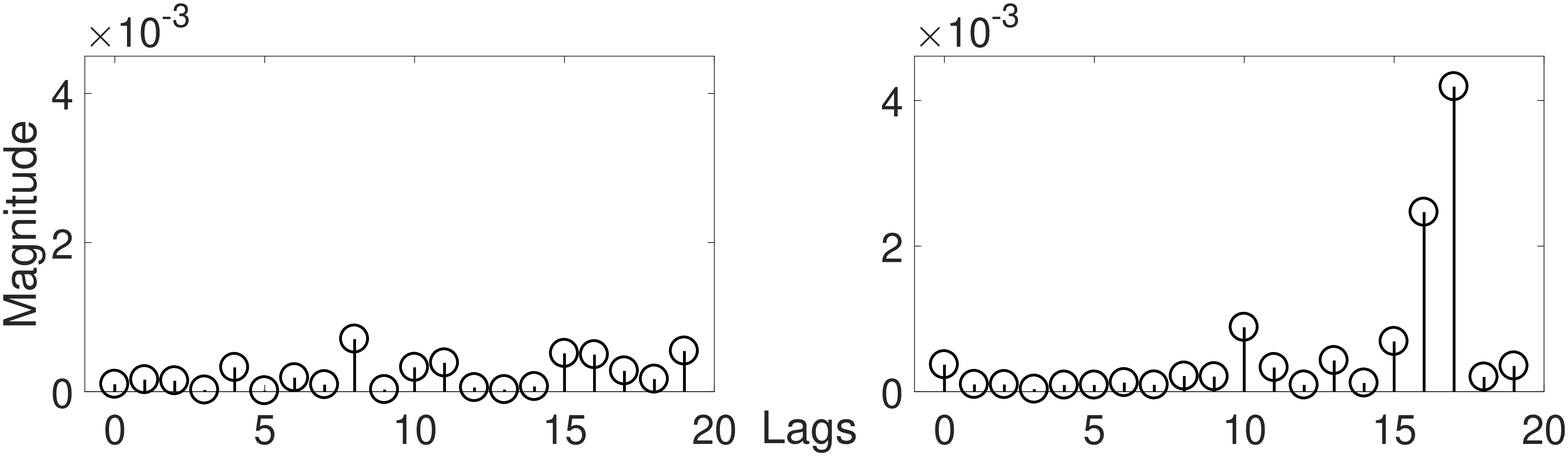}}
%  \vspace{1.5cm}
  \centerline{(e)}\medskip
\end{minipage}
\caption{An example of polynomial dictionary learning methods for the polynomial matrix modeled acoustic impulse denoising, where two polynomial elements are randomly selected to illustrate the denoising performance: Clean FIRs (a);  Noisy FIRs (b);  Reconstructed FIRs by the method in \cite{guan2015polynomial} (c);  Reconstructed FIRs by PMOD+OMP (d);  Reconstructed FIRs by PMOD+POMP (e).}
\label{fig:2}
\end{figure}

In order to show how the proposed polynomial dictionary learning method used for the sparse representation of the signals with time delays, we randomly selected two polynomial elements from the polynomial matrix used to model the acoustic signal in the last experiment. The selected polynomial elements can be seen as a polynomial sub-matrix modeled by two FIRs, which are two segments of the clean test acoustic impulses. Figure \ref{fig:2} shows the clean FIRs in (a), their corresponding noisy FIRs in (b), and the denoised FIRs by the method in \cite{guan2015polynomial} in (c), PMOD+OMP in (d), and PMOD+POMP in (e), respectively. From Figure \ref{fig:2}, we can see that both the proposed method and the method in \cite{guan2015polynomial} are capable of  denoising the acoustic impulses.
%
%%%%
%%
%
% column length use \vfill\pagebreak.
% -------------------------------------------------------------------------
%\vfill
%\pagebreak
%
\section{Conclusions and future work}
\label{sec:Conclusion}

In this paper, we presented a new polynomial dictionary learning method based on the matrix of polynomials model to deal with the signals with time lags, such as acoustic impulse responses. As a byproduct, a polynomial OMP algorithm was also proposed to represent the signals with the learned polynomial dictionary. 
The experiments showed the proposed methods can achieve better performance for acoustic impulse denoising when compared with the baseline methods. 

In our future work, we will further evaluate the two types of polynomial dictionary learning methods, e.g., for the polynomial matrix with different lags. Another potential future work is to extend the current polynomial dictionary learning model, by using a polynomial matrix as the sparse representation coefficient matrix. 

%
%Graphics (i.\,e., illustrations, figures) must not use stipple fill patterns because they will not reproduce properly in Adobe PDF. Please use only SOLID FILL COLORS.
%
%Figures which span 2 columns (i.\,e., occupy full page width) must be placed at the top or bottom of the page.
%
\section{Acknowledgements}
\label{sec:ack}
The work was conducted when J. Guan was visiting the University of Surrey, and supported in part by International Exchange and Cooperation Foundation of Shenzhen City, China (No.  GJHZ20150312114149569). W. Wang was supported in part by the Engineering and Physical Sciences Research Council (EPSRC) Grant Number EP/K014307 and the MOD University Defence Research Collaboration in Signal Processing.

\bibliographystyle{IEEEtran}

\bibliography{polynomial_dict}

\end{document}